\documentclass{elsart}

\usepackage[round,colon,authoryear]{natbib}

\usepackage{epsfig}

\usepackage{amssymb}

\begin{document}

\begin{frontmatter}


% use the thanksref command within \title, \author or \address for footnotes:
% \author{Name\thanksref{label2}}
% including your email address:
% \address{Address\thanksref{email}}
% \thanks[email]{E-mail: }

\title{The Effects of Dust on the Spectral Energy Distributions of
Star-Forming Galaxies}


\author{Daniela Calzetti}

\address{Space Telescope Science Institute, 3700 San Martin Drive, Baltimore, 
MD 21218, U.S.A.; calzetti@stsci.edu}

\begin{abstract}
% Text of abstract
I review the effects of dust, both in absorption and in emission, on
the spectral energy distributions (SEDs) of Local star-forming
galaxies. The energy balance between the stellar light absorbed by
dust at UV-optical-nearIR wavelengths and the energy re-emitted by
dust in the far infrared (FIR) shows that the amount of dust
extinction affecting the galaxy can be predicted with simple
prescriptions applied to the UV-optical data. The implications of
these results for high redshift galaxies are discussed.  Arguments can
be given to support the view that Local starbursts are representative
of high-redshift (z$>$2), UV-detected, star-forming galaxies. If this
is the case, the high redshift FIR emission will be generally
undetected in sub-mm surveys, unless (1) the bolometric luminosity of
the high-z galaxies is comparable to or larger than that of Local
ultraluminous FIR galaxies and (2) their FIR SED contains a cool dust
component.
\end{abstract}


\end{frontmatter}

% main text
\section{Introduction}

\def\putplot#1#2#3#4#5#6#7{\begin{centering} \leavevmode
\vbox to#2{\rule{0pt}{#2}}
\includegraphics{#1}
\end{centering}}
% e.g., \putplot{psfile}{vspace}{angle}{hscale}{vscale}{hoffset}{voffset}
% with vspace in any TeX units, angle in degrees, scale in percent,
% and offset in PostScript points (72/in)
%

A global measure of the impact of dust obscuration on the light
emerging from galaxies is provided by the cosmic background light
spectrum. The background light above $\approx$40~$\mu$m \citep{Hau98,
Fix98} contains between 50\% and $\gtrsim$70\% of the bolometric
emission from galaxies integrated over all redshifts (Figure~1). This
wavelength range is where most of the stellar energy absorbed by dust
is re-emitted.

\begin{figure}[h]
\putplot{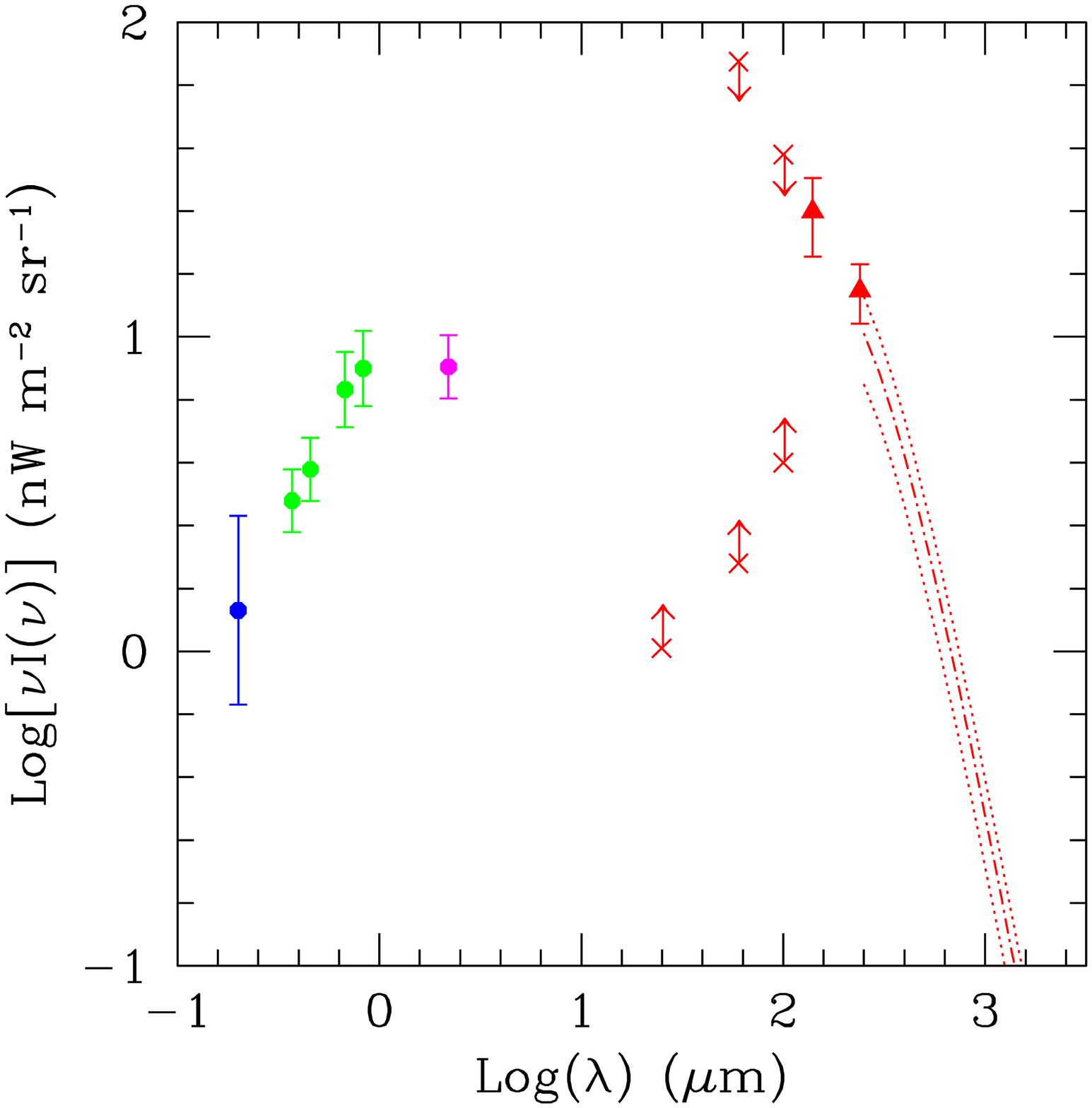}{7.0 cm}{0}{40}{40}{70}{-60}
\caption{The background light emission from the ultraviolet to the far
infrared, given in terms of energy as a function of wavelenght. The UV
data point at 0.2~$\mu$m is from \citet{Arm94}, the optical data are
from the Hubble Deep Field, and the K-band data point from
ground-based surveys \citep{Poz98}. The lower limits in the FIR region
are from IRAS in a no-evolution scenario. The upper limits and the
triangles at $\lambda\ge$60~$\mu$m are from COBE-DIRBE
\citep{Hau98}. The smooth curves are the analytical representation
(dot-dashed line) and the 1~$\sigma$ envelope (dotted lines) of the
background detected by COBE-FIRAS \citep{Fix98}. Adapted from
\citet{Pei99}.}
\end{figure}

While the redshift-integrated effect of dust on galaxy light is
somewhat constrained, the relative importance of dust obscuration as a
function of redshift is not well established, with the possible
exception of the Local Universe. In nearby galaxies, the 8--120~$\mu$m
FIR emission measured by IRAS represents only $\sim$25\% of the
bolometric output from galaxies \citep{Soi91}. However, when compared
with the galaxy UV output, the energy in the IRAS bands represents
between 1/2 and 2/3 of the total UV$+$FIR emission in Local
galaxies. Regions of active star formation emit the bulk of their
energy in the UV, owing to the presence of young, massive stars, and
the dust-reprocessed UV light is the major contributor to the FIR emission
in the IRAS wavebands. Thus, between half and two-third of the star
formation in the Local Universe is obscured by dust.

Observational evidence points to increased dust obscuration in more
distant galaxies \citep{Cal99,Ste99}. Part of this effect is due to
the fact that standard observing techniques access bluer and bluer,
and thus more and more dust-sensitive, restframe wavelength regions
for increasing redshift. Using dust-correction methods tested in the
Local Universe \citep{Cal97,Meu99,Cal00}, the UV light from high
redshift star-forming galaxies has been shown to be possibly obscured
at the 60\%--80\% level. If confirmed by further evidence, this result
has a major impact on a number of properties of the high-z Universe,
including the star formation rate (SFR) estimates
\citep{Ste99}. Moreover, the selective absorption of dust as a
function of wavelength is likely to be the main reason for the
discrepant results given by different SFR indicators. Figure~2 shows 
the SFR density of the Universe as a function of redshift
\citep{Mad96} measured from restframe UV, H$\alpha$, and FIR emission,
before and after dust correction. Only after the effect of dust is
factored in the estimates, the three different SFR indicators are
brought back into agreement.

\begin{figure}
\putplot{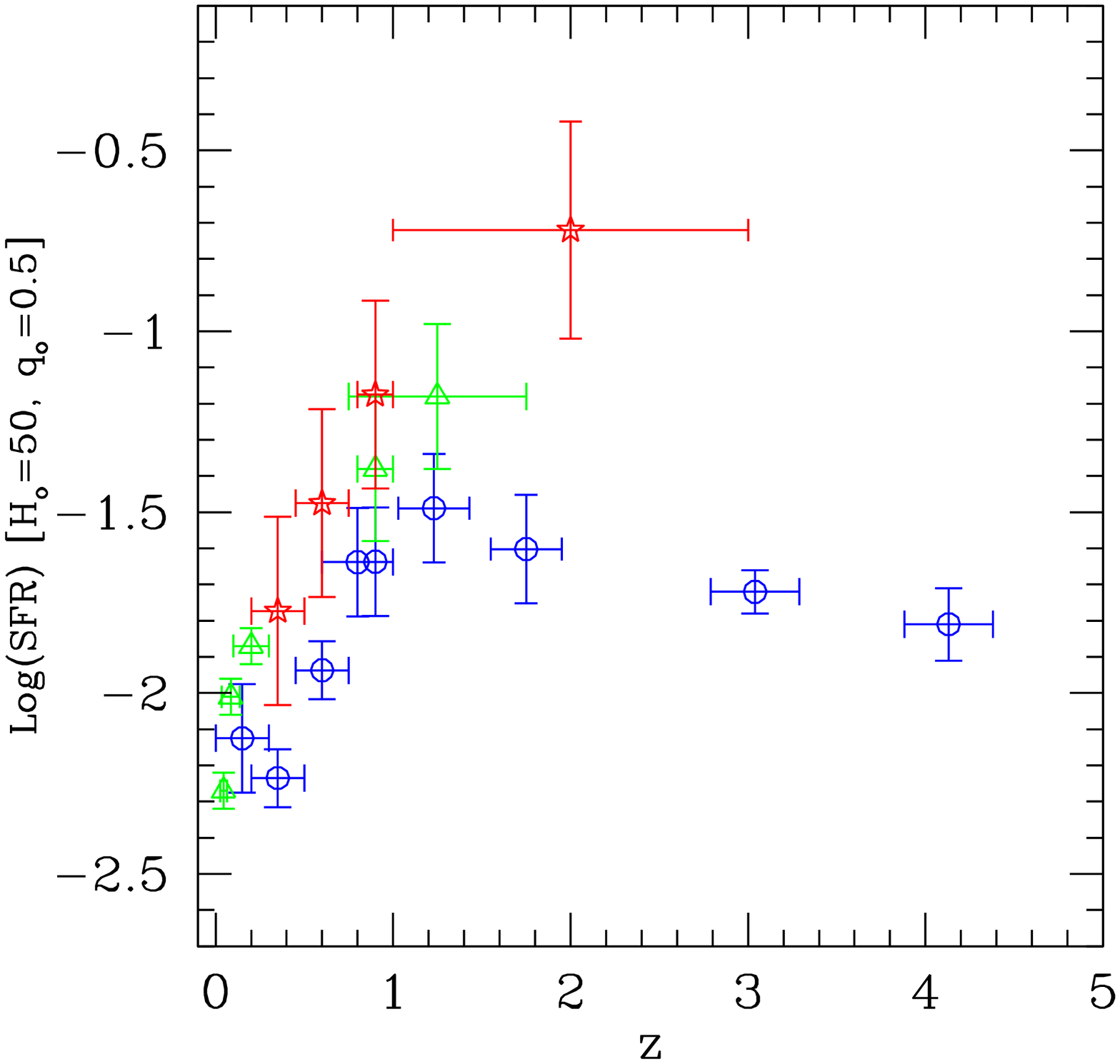}{3.2 cm}{0}{41}{41}{-40}{-83}
\putplot{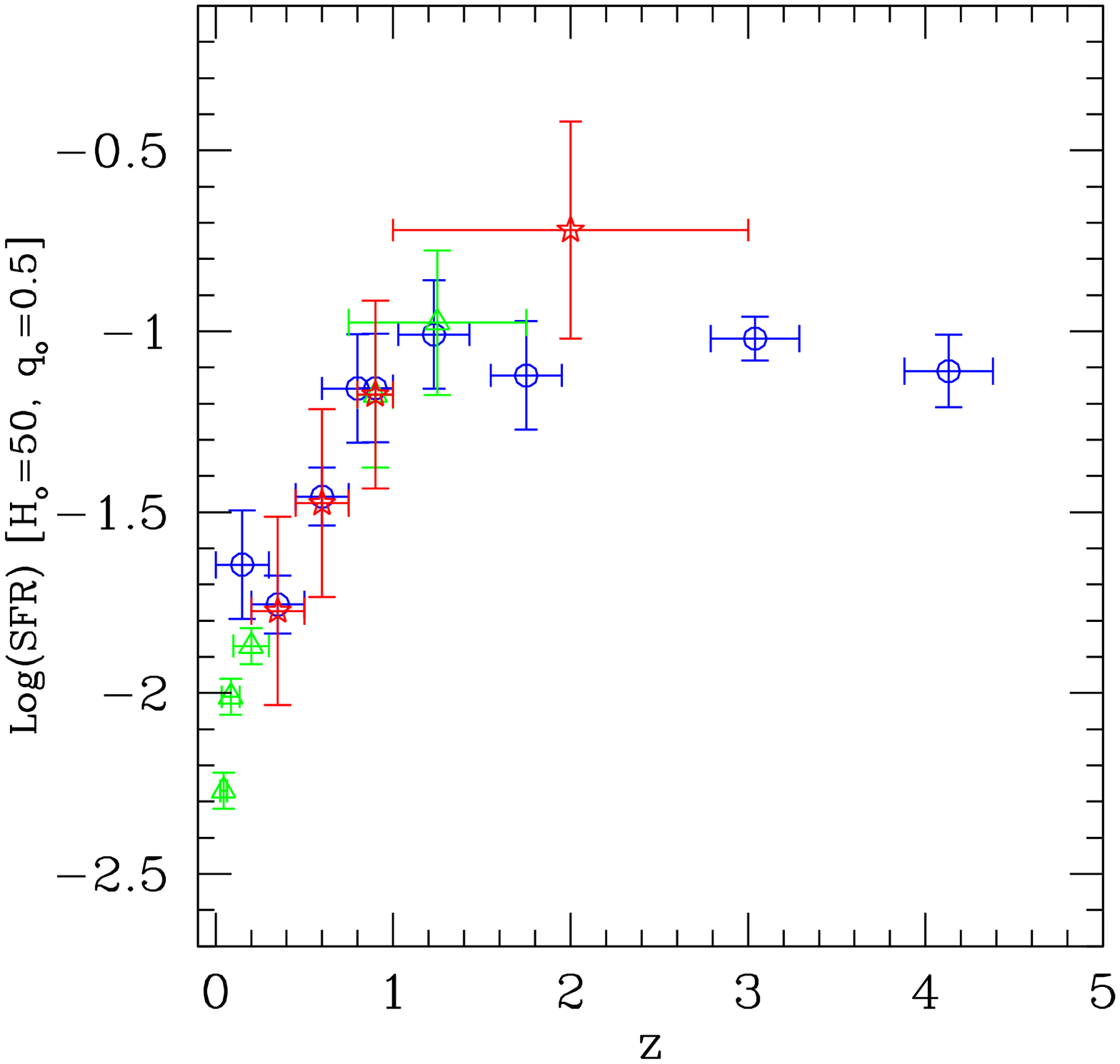}{3.2 cm}{0}{41}{41}{200}{-68} 
\caption{The average star formation rate density of the Universe in
M$_{\odot}$~yr$^{-1}$~Mpc$^{-3}$ as a function of redshift, before
(left panel) and after (rigth panel) correction for the effects of
dust obscuration. Circles indicate restframe UV data at 0.28~$\mu$m
(for z$<$2, \citet{Lil96,Con97}) and at 0.17~$\mu$m (for z$>$2,
\citet{Ste99}). Triangles mark the position of the restframe H$\alpha$
measurements \citep{Gal95,Gron98,Tre98,Gla99,Yan99}. Squares mark the
position of mid/far-IR data, from ISO (z$<$1, \citet{Flo99}) and from
SCUBA \citep{Hug98,Bar00}. 1~$\sigma$ error bars are reported
for all the data.}
\end{figure}

Quantifying the impact of dust on the light emitted by the stellar
component of galaxies bears not only on our understanding of the
evolution of the star formation, but also on a host of other aspects
of galaxy evolution, including the relationship between star
formation and AGN. Various authors \citep{Boy98,Sha99} have pointed
out the close resemblance between the SFR density of galaxies and the
space density of quasars (the latter measured in the radio) as a
function of redshift. The comparison between the two trends is, however, 
strongly dependent on the adopted dust corrections for the
SFR of galaxies, and may or may not result in a similarity. For
instance, corrections like those shown in the right-hand panel of
Figure~2 imply a relatively constant SFR density for z$>$2.5, in a
regime where the space density of quasars is declining for increasing
redshift \citep{Sha99}.

\section{Why Is Dust Elusive?}

Paraphrasing Witt, Thronson \& Capuano (1992), the derivation of the
opacity of a galaxy ``depends sensitively upon the adopted [dust]
geometry, largely due to the important effects of scattering and to
the relative fraction of stars that are lightly obscured''. The
effects of dust obscuration and reddening are very hard to quantify in
galaxies, especially if the only data available are in the
UV-to-nearIR, i.e. in a wavelength range where dust manifests itself
predominantly by absorbing or scattering the stellar light.  Three
main factors conspire to make dust an elusive component of galaxies:
(1) the lack of obvious emission or absorption features below the
K~band (except, in some instances, for the 0.22~$\mu$m~ absorption
`bump'); (2) the grey obscuration often produced by the combination of
dust geometry, varying optical depth, and scattering; (3) the age-dust
degeneracy.

The largest majority of the methods for measuring the opacity/reddening of a
galaxy use the light from the galaxy itself as the 
`illuminating source' for the dust. However, the stellar populations of
galaxies are complex mixtures of stars of all ages. A dust reddened population
often resembles an older population.  The dust-age degeneracy is an important
effect when broad band colors or low resolution spectroscopy only are
available for the object of interest and stellar features are not
identifiable. For instance, an optical attenuation A$_V$=1 changes the galaxy
colors to mimic an age ``increase'' of a factor of $\sim$5 in a stellar
population younger than 100~Myr. Conversely, the colors of a relatively old
stellar population will not be easily discriminated from the colors of a young
but reddened population.

This problem can be minimized in starburst galaxies, where the
UV-to-nearIR output is dominated by the recently formed stars. In
particular, massive stars dominate the UV emission, and their spectra
are in the Rayleigh-Jeans regime above the Lyman discontinuity at
0.0912~$\mu$m. Thus the shape of the UV spectrum of starburst galaxies
is relatively constant over a fairly large range of ages and physical
parameters \citep{Lei95}; the non-ionizing photons are less
age-sensitive than the ionizing photons, and the nebular emission
lines from a starburst will fade before appreciable changes in the
shape of the UV continuum can be observed. In summary, the starburst
populations in galaxies can provide a relatively standard `illuminating
source' for studying the effects of dust in these objects.

\section{Measuring Dust Obscuration in UV-Selected Starbursts}

The ionizing and non-ionizing stellar light absorbed by dust is
re-emitted in the FIR, and measuring dust reddening and obscuration in
galaxies requires multi-wavelength information. Figure~3 shows the
SEDs of two starburst galaxies, covering from the UV
($\sim$0.12~$\mu$m) to the FIR ($\sim$200~$\mu$m). The two SEDs are of
a metal-rich, FIR-luminous galaxy (NGC6090) and a low-metallicity,
Blue Compact Galaxy (Tol1924$-$416). As expected, the metal-rich galaxy
is also dusty, with about 10 times more energy emerging in the FIR
than in the UV--optical range.  In the low-metallicity galaxy, the
energy share between FIR and UV--optical is about 50\% each. Although
the two SEDs are clearly different, the ``action'' of dust on the
stellar light emerging from the two galaxies can be parameterized in a
simple manner, at least in the wavelength range 0.12--2.2~$\mu$m
\citep{Cal94,Cal96,Cal97,Cal00}:
\begin{equation}
F_o(\lambda) = F_i(\lambda)\ 10^{-0.4 A^{\prime}(\lambda)} = F_i(\lambda)\ 10^{-0.4 E_s(B-V)\ k'(\lambda)},
\end{equation}
\begin{equation}
E_s(B-V) = (0.44\pm 0.03) E_g(B-V),
\end{equation}
\begin{eqnarray}
k'(\lambda) &=& 2.659\, (-1.857 + 1.040/\lambda) + 4.05 \ \ \ \ \ \ \ \ \ \ \ \ 0.63\ \mu m \le \lambda \le 2.20\ \mu m \nonumber \\
           &=& 2.659\, (-2.156 + 1.509/\lambda - 0.198/\lambda^2 + 0.011/\lambda^3) + 4.05 \nonumber \\
           & &\ \ \ \ \ \ \ \ \ \ \ \ \ \ \ \ \ \ \ \ \ \ \ \ \ \ \ \ \ \ \ \ \ \ \ \ \ \ \ \ \ \ \ \ \ \ \ 0.12\ \mu m \le \lambda < 0.63\ \mu m.
\end{eqnarray}
\begin{equation}
\beta = 1.9 E_g(B-V) + \beta_o.
\end{equation}
F$_o(\lambda)$ and F$_i(\lambda)$ are the dust-obscured and intrinsic stellar
continuum flux densities, respectively; A$^{\prime}$($\lambda$) is the dust
obscuration suffered by the stellar continuum; E$_s$(B$-$V) and E$_g$(B$-$V)
are the color excess of the stellar continuum and of the nebular emission
lines. The latter is measured using hydrogen line ratios and foreground dust
\citep{Cal96}. $k^{\prime}$($\lambda$) is the starburst obscuration curve, and
$\beta$ is the UV continuum slope measured between 0.125~$\mu$m and
0.26~$\mu$m. $\beta$ is a sensitive indicator of both reddening \citep{Cal94}
and obscuration \citep{Meu99,Cal00}, owing to the fact the the UV spectral
shape is fairly constant when young, massive stars are present.  The value of
the intrinsic UV slope, $\beta_o$, depends slightly on assumptions on the star
formation history of the starburst; $\beta_o$=$-$2.1 and $-$2.3 for constant
star formation over 1~Gyr and over 100~Myr, respectively.

\begin{figure}
\putplot{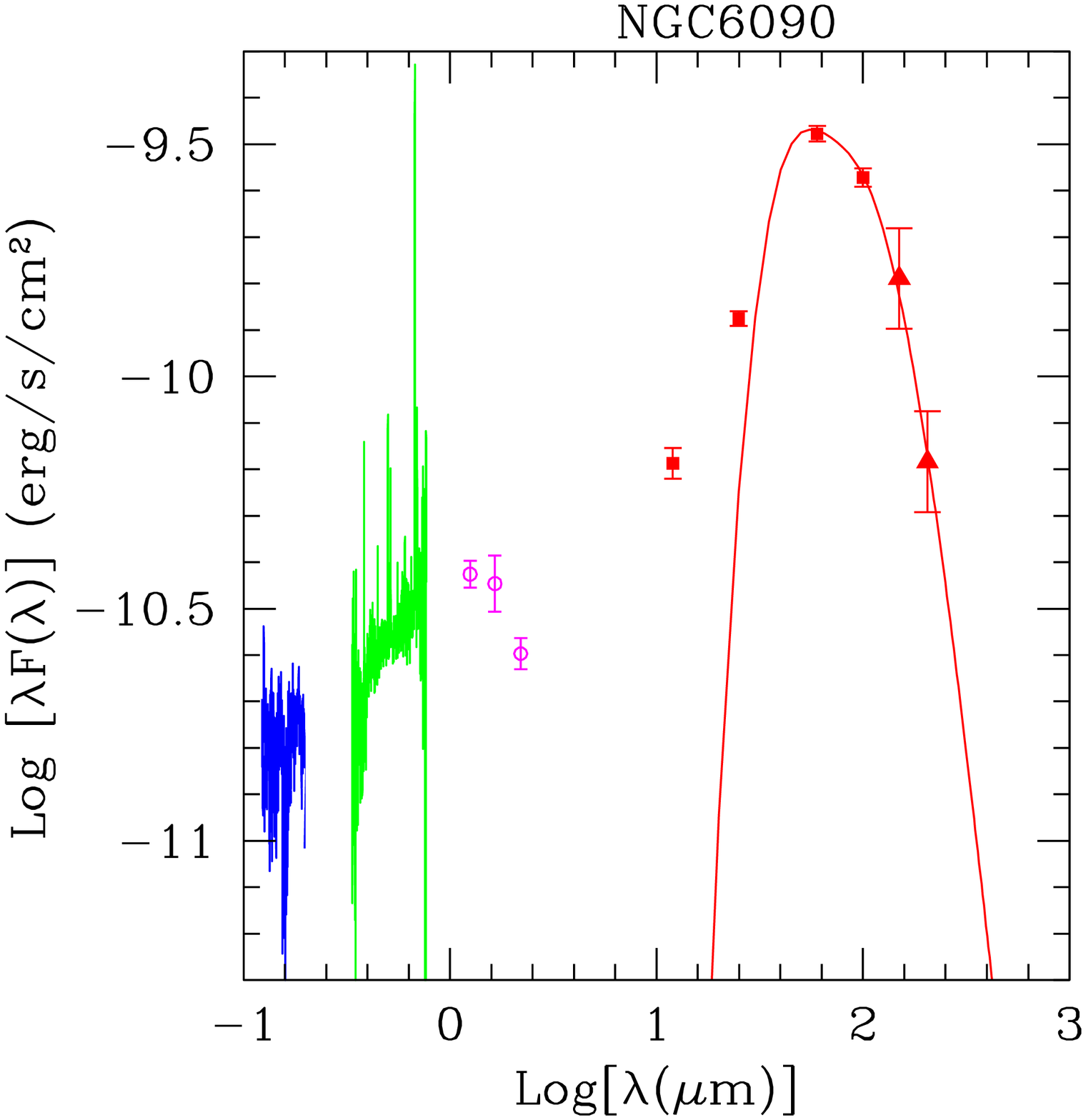}{3.2 cm}{0}{40}{40}{-40}{-73}
\putplot{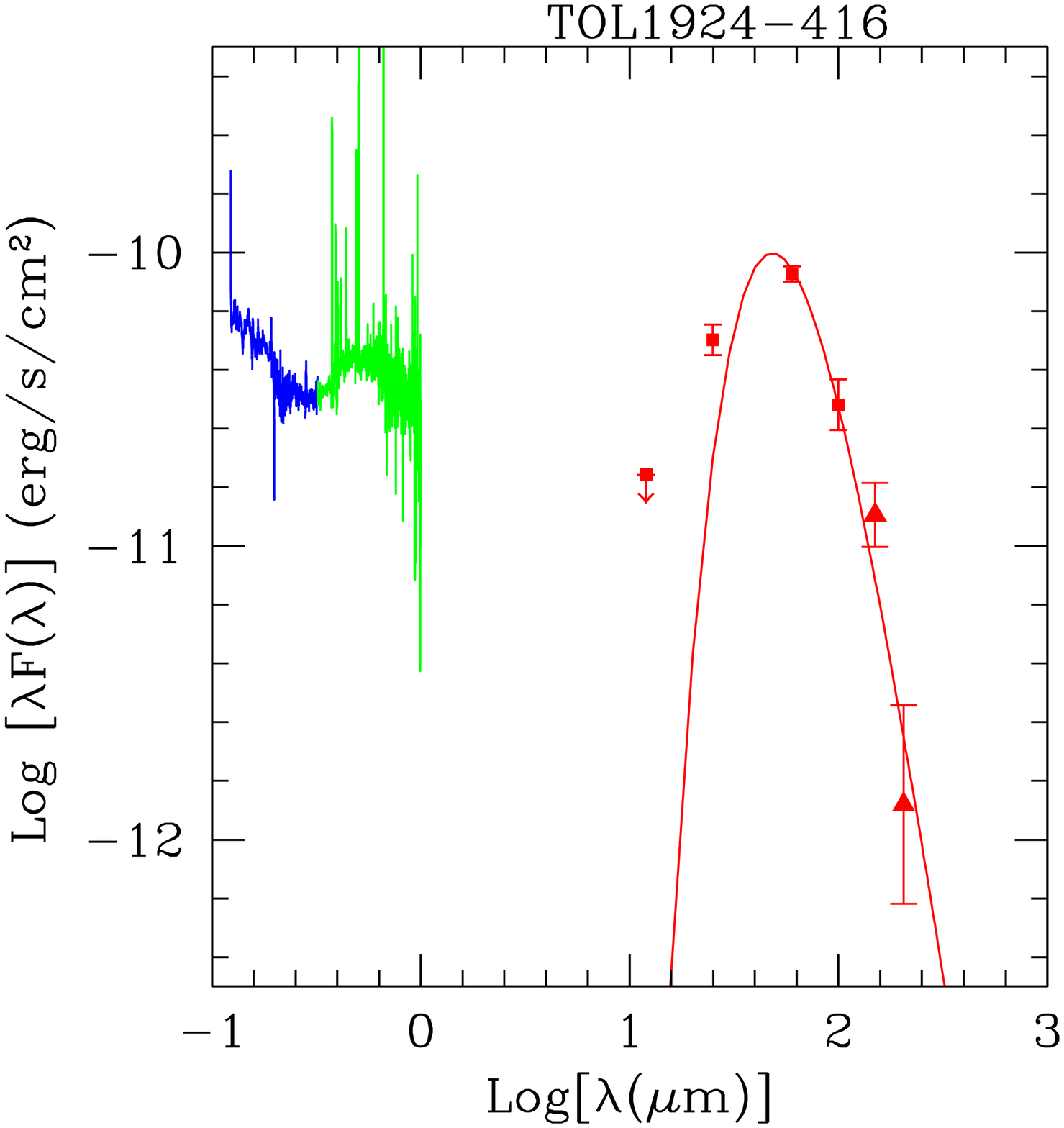}{3.2 cm}{0}{40}{40}{190}{-58} 
\caption{The UV-to-FIR SEDs of two low redshift
starburst galaxies: the FIR--luminous NGC6090 (left panel) and the Blue
Compact Galaxy Tol1924$-$416 (right panel). UV--optical
spectrophotometry is reported as a continuous line in the wavelength
range 0.12-1.0~$\mu$m, nearIR photometry as empty circles, IRAS data
as filled squares, and ISO measurements as filled triangles \citep{Cal00}. 
Upper limits have downward arrows. The smooth curves represent the best 
fit to the data longward of 40~$\mu$m, with a dust emissivity $\propto\nu^2$ 
and multiple temperature components. }
\end{figure}

Equations~1--4 are valid for all starbursts with optical depths less
than a few, typically E(B$-$V)$_g\lesssim$1. In the specific case of
Figure~3, we measure E(B$-$V)$_g$=0.60 in NGC6090 and
E(B$-$V)$_g$=0.02 in Tol1924$-$416. The formulae above can be applied
to the SEDs of starbursts to calculate the amount of ionizing and
non-ionizing stellar energy `lost' to dust absorption, and therefore
predict the FIR dust emission. A comparison with the observed FIR
emission in Local starbursts shows that the agreement between
predictions and observations is within a factor of 2 for each
individual object, and within $\sim$20\% on a sample \citep{Cal00}.
This implies that Equations~1--3 effectively recover the intrinsic
stellar light.  The ratio of the observed fluxes in the 1--1000~$\mu$m
range, F(FIR), and in the ultraviolet at 0.16~$\mu$m, F(UV), is a
measure of the total dust obscuration suffered by the starburst, and
can be expressed in terms of the other dust-obscuration indicator,
$\beta$, as \citep{Meu99,Cal00}:
\begin{equation}
F(FIR)/F(UV)= 1.58 [10^{(0.924 \beta + 1.94)} - 1]
\end{equation}

The wavelength coverage of the FIR SEDs used in this analysis is much
broader than in many previous studies, thanks to the addition of ISO
data; the data span the range $\sim$8--240~$\mu$m (IRAS$+$ISO),
sampling well beyond the peak of the FIR emission and probing the
contribution of dust cooler than $\sim$30~K. It is useful to compare
the flux detected in this broader range with the 40--120~$\mu$m flux,
which has been used to characterize the FIR emission from galaxies
when only IRAS data were available:
\begin{equation}
{F(8-240)\over F(40-120)}= 1.72\pm 0.25.
\end{equation}
For starbursts, the total FIR emission is almost a factor of 2 larger than the
flux detected in the 60~$\mu$m and the 100~$\mu$m IRAS bands, and the flux
beyond 120~$\mu$m represents $\sim$20--25\% of the total \citep{Cal00}.

In addition to providing a simple and purely empirical recipe for dust
obscuration corrections, Equations~1--5 give us insights into the
physical conditions of UV-selected starbursts. Equations~1 and 5 imply
that the dust obscuring the starburst behaves as if it is external to
(surrounding) the region(s) of active star formation. A starburst
environment is likely to be inhospitable to dust, due to the high
energy densities, to the presence of shocks from supernovae
explosions, and to the possibility of gas outflows. Equation~4 means
that the stars ionizing the gas are statistically well mixed with the
stars producing the UV continuum, although Equation~2 implies that the
ionized gas suffers about twice as much reddening as the stars. A way
out from this empasse is to assume that the ionized gas is more
closely associated to the dust than the stars responsible for the UV
emission \citep{Cal94}. The ionizing stars may or may not be as dusty
as the ionized gas \citep{Cal97,Cha00}, while the non-ionizing,
UV-emitting stars will generally be less obscured than the gas
\citep{Cal94}, as they live long enough to diffuse into regions of
lower dust density \citep{Gar82}. HST imaging of nearby starbursts
shows that both scenarios are present, i.e., ionized gas and dust can
be both physically separated from and associated with the exciting
stars \citep{Mai00}. In either scenario, the nebular emission from
the galaxy will be more absorbed than the spatially-integrated stellar
continuum, hence Equation~2.

Equations~1--5 are generally not applicable to the most FIR--luminous
objects, i.e. ULIRGs. In these galaxies, the gas and dust involved in
the starburst are centrally concentrated with densities $\sim$10 times
higher than those of Giant Molecular Clouds \citep{Dow98}. Dust is
heavily mixed with the newly formed stars, and the dominant geometry
is very likely to be more extreme than in UV-selected galaxies. Recent
results from HST show that the application of Equation~5 to ULIRGs
underestimates the FIR flux by factors $\approx$10--100 \citep{Gol00}.

\section{Measuring Dust in UV-Selected High Redshift Galaxies}

The population of Lyman-break galaxies at redshift z$>$2
\citep{Ste96,Ste99} presents a number of observational properties
which resemble the central regions of Local UV-selected starburst
galaxies. The high-redshift objects have typical SFRs per unit area
$\approx$1--2~M$_{\odot}$~yr$^{-1}$~kpc$^{-2}$ \citep{Meu97,Cal99};
the restframe UV spectra show the wealth of absorption lines which is
generally observed when young, massive stars dominate the emission;
nebular emission is also observed in the restframe optical spectra
\citep{Pet98}, confirming that the Lyman-break galaxies are actively
forming new stars (i.e., they are not in a post-starburst
phase). Finally, the range of UV slopes of the high-redshift galaxies
is reminiscent of the one observed in the Local, UV-selected
starbursts, with values $-3\lesssim\beta\lesssim0.4$.

In Local starbursts the large range of UV slopes is a signature of
dust obscuration, and the same interpretation has been suggested for
the high redshift galaxies, based on the similarities presented
above. The median UV slope of the Lyman-break galaxies is
$\beta\sim-$1.4, corresponding to about 80\% of the UV light at
0.17~$\mu$m being reprocessed by dust in the FIR. Incidentally,
$\beta\sim-$1.4 is also the median value of the Local, UV-selected
galaxies; the analogy should not be taken any further as different
selection biases are at work in the two samples. The peak of the FIR
emission from the z$\approx$3 galaxies should be observable at
wavelengths $\approx$200--400~$\mu$m (restframe
$\approx$50--100~$\mu$m). Presently, the most sensitive sub-mm
instrument, SCUBA at the JCMT, reaches the faintest limits at
850~$\mu$m, i.e. restframe wavelength $\approx$200~$\mu$m. This is well
beyond the peak FIR emission observed in galaxies in the Local
Universe (e.g., Figure~3). The FIR energy detected at such long
wavelength represents a fairly small fraction of the total dust
emission; in the specific case of the two starbursts in Figure~3, the
energy emitted at 200~$\mu$m represents between 12\% (NGC6090) and 1.3\%
(Tol1924$-$416) of the total FIR energy. Reconstructing the dust
bolometric emission of high redshift galaxies from SCUBA detections is
thus bound to suffer from large uncertainties, up to a factor
$\approx$10.

If the two galaxies in Figure~3 were placed at z=3, the fluxes
observed at 850~$\mu$m would be 0.92~mJy and $<$0.01~mJy,
respectively, for NGC6090 and Tol1924$-$416. The R band magnitudes
would be 25.7 and 26.4, respectively, about 1.5 mag fainter than the
$\sim$R$^*$ value determined for the Lyman-break galaxies
\citep{Ste99}. The two galaxies differ by less than a factor of 2 in
the restframe UV luminosity, and by about a factor of 400 in the
restframe 200~$\mu$m luminosity. In addition to have a FIR emission
which is about 50 times fainter, Tol1924$-$416 has also a much warmer
dust SED than NGC6090 \citep{Cal00}. This explains why in the
long-wavelength tail of the dust emission the discrepancy in the
intrinsic flux between the two galaxies exharcerbates.

If the luminosity of a z=3 galaxy with the SED of NGC6090 is boosted
up to be detectable by SCUBA (adopting as fiducial value the
5~$\sigma$ limit of 2~mJy in the Hubble Deep Field, \citet{Hug98}),
the galaxy would have R$\sim$24.9 and a bolometric luminosity
L$_{bol}$=1.5$\times$10$^{12}$~L$_{\odot}$. Thus, such a galaxy would
need to be as luminous as a ULIRG to be observed with SCUBA
\citep{Lil99}. A more extreme situation is seen with Tol1924$-$416,
due to the very small amount of energy emitted at 200~$\mu$m, less
than 1\% of the bolometric luminosity. To be detectable with SCUBA, a
z=3 galaxy with the SED of Tol1924$-$416 would need to have a
bolometric luminosity L$_{bol}\sim$3$\times$10$^{13}$~L$_{\odot}$, a
few times brighter than the brightest non-AGN powered galaxy known in
the Local Universe. Such a galaxy would have R$\sim$19, almost 4~mag
brighter than the brightest Lyman-break galaxy identified so far
(Giavalisco, 2000, private communication). Clearly, a galaxy with a
luminosity within observed limits and with a FIR SED as warm as that
of Tol1924$-$416 would not be detectable with SCUBA.

To summarize, a z=3 galaxy with an SED similar to that of Local
UV-selected starbursts needs to be dusty, as bright as a ULIRG, and to
have a relatively cool FIR SED to be detectable with current sub-mm
instrumentation. The first two criteria already restrict the number of
candidates, since, even if high redshift galaxies may be dusty, it is
not established how many of those have bolometric luminosities in the
10$^{12}$~L$_{\odot}$ range. The third criterion (cool FIR SED) is
equivalent to require that, for an emissivity index $\epsilon$=2, dust
with temperature $\approx$20~K must be present in the galaxy; this
criterion may further bring down the number of candidates, depending
on the details of the distribution of dust and star formation within
the galaxy. A recent project aimed at observing Lyman-break galaxies
with SCUBA has yielded only one tentative detection out of $\sim$10
targets \citep{Cha99}, confirming that a number of factors have to
play together to make the long wavelength tail of the FIR emission of
those galaxies detectable. Photometric uncertainties in the restframe
UV measurements and Malmquist bias add to the difficulty of making
reliable predictions for the FIR luminosity of the high-z galaxies
\citep{Ade00}.

The effects of dust in the distant Universe still need to be placed on
firmer observational ground. To achieve this target, high redshift
star-forming galaxies require better spectral energy coverage. The
available optical observations will need to be supplemented with near- to
mid-IR spectral data to measure the restframe optical emission and
nebular lines, and with far-IR/sub-mm/mm data to derive a full picture
of the dust emission. SIRTF, ALMA, and NGST are the key missions which
will address this issue. Knowing how much dust is present in galaxies
at different redshifts will not only allow us to derive the intrinsic
star formation history of the Universe, but will also shed light on a
number of other astrophysical issues, such as the dust and metal
enrichment of the evolving galaxies and of the Universe, and the
starburst--AGN connection.

\section{Acknowledgements}

I would like to thank the Organizing Committee of the FIRSED~2000
Workshop, and in particular Peter Barthel, for the invitation to this
stimulating meeting, for financially supporting my trip, and for
arranging for a full-time baby sitter for my 4-months old son. I also
thank Claus Leitherer and Gerhardt Meurer for a critical reading of this
manuscript.
\label{}

% Bibliographic references with the natbib package:
% Parenthetical: \citep{Bai92} produces (Bailyn 1992).
% Textual: \citet{Bai95} produces Bailyn et al. (1995).
% An affix and part of a reference:
%   \citep[e.g.][Ch. 2]{Bar76}
%   produces (e.g. Barnes et al. 1976, Ch. 2).

\end{document}